\documentstyle{aipproc}

\begin{document}
\title{Asymptotic Behavior of the Wave Packet Propagation through a Barrier : \\
  the Green's Function Approach Revisited}

\author{Bogdan~Mihaila\footnote{Present address:
Physics Division, Argonne National Laboratory, Argonne, IL 60439;
}
    \thanks{electronic mail:bogdan@theory.phy.anl.gov}
$^{a,b}$,
       Shmuel~A.~Gurvitz$^{c,d}$,
    David~Dean$^{a,e}$,
        Witold~Nazarewicz$^{a,e,f}$
}

\address{
  $^a$ Physics Division,
      Oak Ridge National Laboratory,
      P.O. Box 2008, Oak Ridge, TN 37831 \\
  $^a$ Physics Division,
      Oak Ridge National Laboratory,
      P.O. Box 2008, Oak Ridge, TN 37831 \\
  $^b$ Chemistry and Physics Department,
      Coastal Carolina University,
      Conway, SC 29528-6054 \\
 $^c$ Department of Physics,
      The Weizmann Institute of Science,
      Rehovot 76100, Israel \\
 $^d$ Joint Institute for Heavy Ion Research,
      Oak Ridge, Tennessee 37831 \\
 $^e$ Department of Physics,
      University of Tennessee,
      Knoxville, TN 37996 \\
 $^f$ Institute of Theoretical Physics,
      Warsaw University,
      Ho\.za 69, PL-00681, Warsaw, Poland
}

\maketitle

\begin{abstract}
  To model the decay of a quasibound state we use the modified
  two-potential approach introduced by Gurvitz and
  Kalbermann~\cite{ref:gurvitz_a,ref:gurvitz_b}.  This method has
  proved itself useful in the past for calculating the decay width and
  the energy shift of an isolated quasistationary
  state~\cite{ref:witek_pe}.  We follow the same approach in order to
  propagate the wave-packet in time with the ultimate goal of
  extracting the momentum-distribution of emitted particles.  The
  advantage of the method is that it provides the time-dependent wave
  function in a simple semi-analytic form.  We intend to apply this
  method to the modeling of metastable states for which no direct
  integration of the time-dependent Schr\"odinger equation is
  available today.
\end{abstract}

The Two Potential Approximation (TPA) introduced in
Refs.~\cite{ref:gurvitz_a,ref:gurvitz_b} turned out to be an extremely
successful tool for the description of a metastable state. Simple
expressions based on the TPA made it possible to obtain a very precise
estimate of the life-time of a very narrow resonance without the need
of introducing an explicit time dependence. In this work, we use the
TPA in order to derive the equations describing the time evolution of
the wave function of a particle tunneling through a
spherically-symmetric barrier.

Let us consider a particle moving in a central potential $V(r)$ with a
barrier. Asymptotically, i.e., at large values of $r$, we assume that
$V(r)$$\rightarrow$0.  In the TPA, $V(r)$ can be decomposed as
\begin{equation}\label{Vofr}
 V(r) \ = \ U(r) \ + \ W(r) \>,
\end{equation}
\newpage
\noindent
where
\begin{equation}\label{Uofr}
U(r)=\left\{ \begin{array}{ll} V(r) & \mbox{if $r<R$} \\ V(R) &
\mbox{if $r>R$}
\end{array}
\right.
\end{equation}
is an auxiliary potential that produces a bound state at energy $E_0$
close to the energy of the metastable state, and $W(r)$ is a
``closing" potential which is treated perturbatively.  The separation
radius $R$ should be chosen far from the classical turning
points~\cite{gns}.

At $t$=0 the initial state is taken to be the bound eigenstate
$\Phi_0(\vec r)$ of the auxiliary Hamiltonian
\begin{equation}
H_0 \ = \ T \ + \ U(r)
\end{equation}
(we take $\hbar$=1).  In the following we assume that $\Phi_0(\vec r)$
is well isolated, i.e., it is well separated from the remaining bound
states of $U(r)$ having the same quantum numbers.  In such a case, at
$t$$>$0, the wave packet represented by the wave function $\Psi(\vec
r, t)$ can be expanded in the basis $\{\Phi_0(\vec r), \Phi_k(\vec{r})
\}$:
\begin{eqnarray}
   \Psi(\vec r, t) & = & b_0(t) \Phi_0(\vec r) e^{- i E_0 t}
   \ + \ \int \frac{d^3 k}{(2 \pi)^3} \ b_k(t) \Phi_k(\vec{r}) e^{-i
   E_k t} \>,
\label{eq:wave_packet}
\end{eqnarray}
with the initial conditions $b_0(t=0)$=1 and $b_k(t=0)$=0.  In
Eq.~(\ref{eq:wave_packet}) the wave functions $ \Phi_k(\vec{r})$
represent the continuum and $E_k=V(R)+k^2/2m$.  We shall refer to the
first and second terms above, as $\Psi_{I}(\vec r, t)$ and
$\Psi_{II}(\vec r, t)$, respectively.

To evaluate the two components, $\Psi_{I}(\vec r, t)$ and
$\Psi_{II}(\vec r, t)$, the Laplace transform method can be applied.
In terms of the Laplace-transformed expansion coefficients~$b(t)$,
\begin{eqnarray}
   \tilde b(\varepsilon) \ = \ \int_0^\infty b(t) e^{i \varepsilon t}
   \ dt,
\end{eqnarray}
the Laplace transform of the wave packet $\Psi(\vec r, t)$ can be
written as
\begin{eqnarray}
   \Psi_{I}(\vec r, \varepsilon + E_0) & = & \tilde b_0(\varepsilon)
   \, \Phi_0(\vec r) \>,
   \label{eq:PsiI_til}
   \\ \Psi_{II}(\vec r, \varepsilon + E_0) & = & \int \frac{d^3 k}{(2
   \pi)^3} \ \tilde{\bar b}_k(\varepsilon_k) \, \Phi_k(\vec{r}) \>,
   \label{eq:PsiII_til}
\end{eqnarray}
where $\bar b_k(t) = e^{-i V(R) t} b_k(t)$ and $\varepsilon_k =
\varepsilon + E_0 + V(R)- E_k$.

Assuming a spherically symmetric potential~$V(r)$, the coefficient
$\tilde b_0(\varepsilon)$ has been calculated as \cite{ref:gurvitz_b}
\begin{equation}
   \tilde b_0(\varepsilon) \ = \ \frac{i}{ \varepsilon - \varepsilon_0
   } \>,
\end{equation}
\newpage
\noindent
with
\begin{eqnarray}
   \varepsilon_0 & = & \Delta - i \, \frac{\Gamma}{2} \\ & = & - \
   \sqrt{\frac{\pi}{2}} \ \frac{|\phi_0(R)|^2}{2m k_0} \ [ \alpha
   \chi_{lk_0}(R) + \chi_{lk_0}'(R) ] \
   [ \alpha \chi^{(+)}_{lk_0}(R) + \chi^{(+)}_{lk_0} {}'(R) ] \>,
   \nonumber
\end{eqnarray}
where $\alpha = \sqrt{2m(V_0-E_0)}$ and
$k_0=\sqrt{2m(E_0+\varepsilon_0)}$.  In this work, $\phi_0(r)$ is the
radial wave function of $\Psi_0$ and $\chi_{lk}(r)$ and
$\chi^{(+)}_{lk}(r)$ are, respectively, the regular and outgoing waves
of the Hamiltonian with the potential $\tilde{W}(r) = W(r) + V(R)$.
(Note, that our radial continuum functions satisfy the orthogonality
and completeness relationships
\begin{eqnarray}
   \int_0^\infty \ \chi_{lk}^*(r) \, \chi_{lk'}(r) \ dr \ = \
   \delta(k-k') \>, \\ \int_0^\infty \ \chi_{lk}^*(r) \, \chi_{lk}(r')
   \ dk \ = \ \delta(r-r') \>.
\end{eqnarray}
Compared with expressions in Refs.~\cite{ref:gurvitz_a,ref:gurvitz_b},
this results in an additional factor of $\sqrt{\pi/2}$ in the front of
every $\chi_{lk}(r) function$~\cite{ref:Baz}.)

With the above definitions, the radial part of the first component in
Eq.~(\ref{eq:wave_packet}) is
\begin{equation}
   \psi_{I}(r, t) \ = \ \frac{\phi_0(r)}{r} \ e^{-i
   (E_0+\varepsilon_0) t} \>.
\end{equation}
The coefficients $\tilde {\bar b}_k(\varepsilon_k)$ are determined by
solving the system of integral equations
\begin{eqnarray}\label{setb}
   \varepsilon \tilde b_0(\varepsilon) & = & i + W_{00} \tilde
   b_0(\varepsilon) + \int \frac{d^3 k}{(2 \pi)^3} \ \tilde{W}_{0k}
   \tilde{\bar b}_k(\varepsilon_k) \>, \\ \nonumber \varepsilon_k
   \tilde{\bar b}_k(\varepsilon_k) & = & W_{k0} \tilde
   b_0(\varepsilon) + \int \frac{d^3 k'}{(2 \pi)^3} \ \tilde{W}_{kk'}
   \tilde{\bar b}_{k'}(\varepsilon_{k'}) \>,
\end{eqnarray}
with $\tilde{W}_{kk'}\equiv\langle\Phi_k|\tilde{W}|\Phi_{k'}\rangle$.
The solution of (\ref{setb}) can be formally written as
\begin{eqnarray*}
   \tilde{\bar b}_k(\varepsilon_k) & = & \frac{1}{\varepsilon_k} \
   \langle \Phi_k \, | \, \Bigl ( 1 \ + \ \tilde{W} \tilde G_0 \ + \
   \tilde{W} \tilde G_0 \tilde{W} \tilde G_0 \ + \ \cdots \, \Bigr )
   \, W \, | \, \Phi_0 \rangle \ \tilde b_0(\varepsilon) \>,
\end{eqnarray*}
where
\begin{eqnarray}
   \tilde G_0 \ = \ \int \frac{d^3 k}{(2 \pi)^3} \ \frac{ | \Phi_k
   \rangle \, \langle \Phi_k | }{\varepsilon_k} \>.
\end{eqnarray}

The outgoing part of the wave function, $\Psi_{II}(\vec r,
\varepsilon)$, can now be expressed in terms of $\Psi_{I}(\vec r,
\varepsilon)$ as
\begin{eqnarray}
   \tilde \Psi_{II}(\vec r, \varepsilon+E_0) & = & \int d^3 r' \
   \tilde G(\varepsilon+E_0;\vec r',\vec r') \ W(r') \ \tilde
   \Psi_{I}(\vec r, \varepsilon+E_0) \>,
\end{eqnarray}
where we now introduce the Green's function
\begin{equation}
   \tilde G(E) = \tilde G_0(E) + \tilde G_0(E) \, \tilde{W} \, \tilde
   G(E) \ = \ (1 - \Lambda) (E - H + \Lambda \tilde{W})^{-1},
\end{equation}
with $\Lambda= | \Phi_0 \rangle \, \langle \Phi_0 |$ being the
projection operator on $\Phi_0$.  The Green's function $\tilde G(E)$
is approximated in the spirit of~Ref.\cite{ref:gurvitz_b} by
neglecting the contribution from~$\Lambda$, and then by replacing the
potential $V(r)$ by $\tilde{W}(r)$. This gives $\tilde{G}(E)\approx
G_{\tilde{W}}(E)$, where
\begin{equation}
G_{\tilde{W}}(E) = (E-H_{\tilde{W}})^{-1}
\end{equation}
is the Green's function of $H_{\tilde{W}}=T+\tilde{W}$.

By taking the inverse Laplace transform of (\ref{eq:PsiII_til}), one
obtains for the the radial wave function
\begin{equation}\label{wf2}
 \psi_{II}(r, t) = \frac{1}{2\pi} \ \int_R^\infty \ r' dr' \ W(r')
   \phi_0(r') \int_{i \gamma - \infty}^{i \gamma + \infty} \
   d\varepsilon \ e^{-i \varepsilon t} G_{\tilde{W}}(\varepsilon;r,r')
   \tilde b_0(\varepsilon-E_0) \>.
\end{equation}
The $\varepsilon$-integral is evaluated using the residue theorem, and
results in the sum of the residues corresponding to the two poles of
the integrand.

\subsection*{Contribution due to the pole of \mbox{
\boldmath ${\tilde b}_0(\varepsilon-E_0)$}}

Using the standard techniques explained in Ref.~\cite{ref:gurvitz_b},
we obtain
\begin{equation}
   \psi_{II,a}(r<R, t) = \sqrt{\frac{\pi}{2}} \ \frac{\phi_0(R)}{k_0
   r} \ [ \alpha \chi_{l k_0}^{(+)}(R) + \chi_{l k_0}^{(+)} {}'(R) ] \
   \chi_{l k_0}(r) \ e^{-i (E_0 + \varepsilon_0) t} \>,
\end{equation}
and
\begin{eqnarray}
   \psi_{II,a}(r>R, t) & = & - \, \frac{\phi_0(r)}{r} \ e^{-i (E_0 +
   \varepsilon_0) t} \nonumber \\ && \ + \ \sqrt{ \frac{\pi}{2}}
   \frac{\phi_0(R)}{k_0 r} [ \alpha \chi_{l k_0}(R) + \chi_{l k_0}'(R)
   ] \chi_{l k_0}^{(+)}(r) e^{-i (E_0 + \varepsilon_0) t}
   \>.\label{part2}
  \end{eqnarray}
Note that for $r>R$ the contribution from $\psi_I$ is exactly canceled
by the first term in (\ref{part2}).

\subsection*{Contribution due to the pole of the Green's function.}

The Green's function $G_{\tilde{W}}$ has a continuum of simple poles
along the real $E>0$ axis. After using the spectral representation of
$G_{\bar W}(\varepsilon;r,r')$, one can express $\psi_{II,b}(r,t)$ as
\begin{equation}\label{ooo}
   \psi_{II,b}(r, t) = \frac{2m}{r} \ \int_R^\infty \ dr' \ W(r') \
    \phi_0(r') \
\label{eq:psi_IIres}
   \int_0^\infty \ dk \ \frac{e^{-i \frac{k^2}{2m} t}} {k^2 - k_0^2} \
       \chi_{lk}(r) \, \chi_{lk}^*(r') \>.
\end{equation}
The evaluation of the integral~(\ref{eq:psi_IIres}) represents the
corner stone of the present approach.

For now we will restrict ourselves to making some remarks regarding
the asymptotic behavior of this integral at large values of $r$.  We
shall also assume that the potential $V(r)$ has finite range (i.e., it
vanishes at large values of $r$). While this assumption cannot be used
for the case of the Coulomb potential, it is still interesting to
investigate the general structure of the solution for the short-range
potential.  In this limit, the $S$-matrix is meromorphic in the
complete complex~$k$-plane, and~\cite{ref:Baz,Newton}
\begin{equation}\label{Smat}
[S_l(k)]^* = S_l(-k^*),~~{\rm and}~~S_l(k)=[S_l(-k)]^{-1}.
\end{equation}
Expressing $\chi_{lk}(r)$ in the asymptotic form:
\begin{equation}
\chi_{lk}(r)=\sqrt{\frac{2}{\pi}}\frac{1}{2i}\left(S^{1/2}_l(k)e^{ikr}
-(-)^lS^{-1/2}_l(k)e^{-ikr}\right),
\end{equation}
one obtains for the $k$-integral in (\ref{ooo}):
\begin{eqnarray}
 I(r,r',t) &=& \int_0^\infty \ dk \ \frac{e^{-i \frac{k^2}{2m} t}}
   {k^2 - k_0^2} \ \chi_{lk}(r) \, \chi_{lk}^*(r')
   \label{eq:partII}
   \\ && \ \asymp \ \frac{1}{2\pi} \ \int_{-\infty}^\infty \ dk \
   \frac{e^{-i \frac{k^2}{2m} t}} {k^2 - k_0^2} \ \left [ e^{i k
   (r-r')} \ - \ (-)^l \, S_l(k) \, e^{i k (r+r')} \right ] \>.
\label{eq:ass_int}
\end{eqnarray}

The integrand in Eq.~(\ref{eq:ass_int}) is a sum involving two complex
functions of complex~$k$,
\begin{eqnarray}
   \frac{1}{k^2 - k_0^2} \>, \qquad {\rm and} \qquad \frac{ S_l(k)
   }{k^2 - k_0^2} \>,
\label{eq:integrand}
\end{eqnarray}
which have common poles at $\pm k_0$. In addition, $S_l(k)$ has an
infinite number of simple poles. They are located in the lower half of
the complex $k$-plane, symmetrically with respect to the imaginary
axis. Following the notation of van Dijk and Nogami~\cite{ref:mosh},
we shall denote the poles in the fourth quadrant with $k_\nu,
\nu=1,2,3,\ldots$, and the poles in the third quadrant with $k_\nu,
\nu=-1,-2,-3,\ldots$ It follows from Eq.~(\ref{Smat}) that
\begin{eqnarray}
   {\rm Re}(k_\nu) \ = \ -\, {\rm Re}(k_{-\nu}) \>, \qquad \qquad {\rm
   Im}(k_\nu) \ = \ {\rm Im}(k_{-\nu}) \>.
\end{eqnarray}
In the following, the residue of the $S_l(k)$ at the pole $k_\nu$ is
denoted by $b_\nu$.

Since the complex function~(\ref{eq:integrand}) has no essential
singularity at infinity, we can apply the Mittag-Leffler theorem in
order to obtain a pole expansion for~(\ref{eq:integrand}).
Consequently, Eq.~(\ref{eq:integrand}) can be replaced by
\begin{eqnarray}
   && \frac{1}{2k_0} \,\left ( \frac{1}{k - k_0}-\frac{1}{k +
   k_0}\right ) \>,
\label{eq:subst_1}
   \qquad {\rm and} \\ && \frac{S_l(k_0)}{2k_0} \, \frac{1}{k - k_0}
 -\frac{S_l(-k_0)}{2k_0} \, \frac{1}{k + k_0} \ + \ \sum_{\nu=-\infty}^\infty \ \frac{b_\nu}{k_\nu^2 - k_0^2} \ \frac{1}{k - k_\nu} \>.
\label{eq:subst_2}
\end{eqnarray}
By substituting Eqs.~(\ref{eq:subst_1}) and (\ref{eq:subst_2}) in
(\ref{eq:ass_int}), the integral~(\ref{eq:ass_int}) becomes
\begin{eqnarray}
   I(r,r',t) & = &
   \int_{-\infty}^\infty \ dk \
        \frac{e^{-i \frac{k^2}{2m} t} e^{i k (r-r')} }
             {4\pi k_0}\left ({1\over k-k_0}-{1\over k+k_0}\right )
\label{eq:almost}
   \\ &&
   - \frac{(-)^l}{2\pi}
   \Biggl [
   \int_{-\infty}^\infty dk
        \frac{e^{-i \frac{k^2}{2m} t} e^{i k (r+r')} }
             {2k_0}\left ({S_l(k_0)\over k-k_0}-{S_l(-k_0)\over k+k_0}
   \right )
   \nonumber \\ && \hspace{1in}
   +
   \sum_{\nu=-\infty}^\infty \frac{b_\nu}{k_\nu^2 - k_0^2}
   \int_{-\infty}^\infty dk
       \frac{e^{-i \frac{k^2}{2m} t} e^{i k (r+r')} }{k - k_\nu}
   \Biggr ]
   \>.
   \nonumber
\end{eqnarray}
The above can be now expressed in terms of the Moshinsky function
\begin{eqnarray}
   M \left ( k,{\cal R},\tau \right )
   & = & \frac{i}{2\pi} \
   \int_{-\infty}^\infty \ dp \
       \frac{e^{-i p^2 \tau} \, e^{-i \, p {\cal R}} }
            {p - k}
   \nonumber \\ & = &
   \frac{1}{2} \ e^{-i k^2 \tau} \, e^{-i \, k {\cal R}} \
   {\rm erfc}(y)
   \>,
\end{eqnarray}
where
\[ y \ = \ e^{-i\pi / 4} \ \sqrt{\tau} \
           \left ( \frac{\cal R}{2 \tau} - k \right ) \>,
\]
where $\tau=t/2m$ and ${\cal R}=r\pm r'$.
The integral $I(r,r',t)$ can now be calculated in the closed form:
\begin{eqnarray}
   I(r,r',t) & = &
   \frac{1}{2i k_0}
   \Biggl [ M \left ( k_0, r-r', \frac{t}{2m} \right )
   + M \left ( k_0, r'-r, \frac{t}{2m} \right )
   \nonumber \\ &&
   - (-)^l S_l(k_0)
            M \left ( k_0, r+r', \frac{t}{2m} \right )
   - (-)^l S_l(k_0)
            M \left ( k_0, -r-r', \frac{t}{2m} \right )
   \Biggr ]
   \nonumber \\ &&
   + \ i \, (-)^l
   \sum_{\nu=-\infty}^\infty \frac{b_\nu}{k_\nu^2 - k_0^2}
       M \left ( k_\nu, r+r', \frac{t}{2m} \right )
   \>.
\label{eq:as_final}
\end{eqnarray}
This concludes our derivation.

\vskip 0.5in

\centerline{\bf\Large Acknowledgments} \vskip 0.4cm This
work was supported by the U.S. Department of Energy under Contract
Nos.\ DE-FG02-96ER40963 (University of Tennessee), DE-FG05-87ER40361
(Joint Institute for Heavy Ion Research), and DE-AC05-96OR22464 with
Lockheed Martin Energy Research Corp.\ (Oak Ridge National
Laboratory).

\end{document}